\documentclass[twocolumn,showpacs,preprintnumbers,amsmath,amssymb,showkeys,aps,prl]{revtex4}
\usepackage{dcolumn}
\usepackage{bm}
\input{graphicx}
\begin{document}

\newcommand{\nit}{$^{14}{\rm N}$}
\newcommand{\ox}{$^{15}{\rm O}$}
\newcommand{\ecm}{E$_{\rm cm}$}
\newcommand{\ep}{E$_{p}^{lab}$}
\newcommand{\ex}{E$_{\rm x}$}
\newcommand{\tnine}{T$_{9}$}
\newcommand{\wg}{$\omega\gamma$}
\newcommand{\hed}{$(^3{\rm He},d)$}
\newcommand{\he}{$^3{\rm He}$}
\newcommand{\npg}{$^{14}{\rm N}(p,\gamma)^{15}{\rm O}$}
\newcommand{\pg}{(p,$\gamma$)}
\newcommand{\nhed}{$^{14}{\rm N}(^3{\rm He},d)^{15}{\rm O}$}

\title{Direct measurement of the $^{14}$N(p,$\gamma$)$^{15}$O $S$-factor}

\author{R.~C.~Runkle$^{1,2}$, A.~E.~Champagne$^{1,2}$, C.~Angulo$^{3}$, C.~Fox$^{1,2}$, C.~Iliadis$^{1,2}$, R.~Longland$^{1,2}$, and J.~ Pollanen$^{1,2}$ }

\affiliation{$^{1}$The  University of North Carolina at Chapel Hill, Chapel 
Hill, North Carolina 27599-3255, USA\\
$^{2}$Triangle Universities Nuclear Laboratory, Durham, North Carolina, 
27708-0308, USA\\
$^{3}$Centre de Recherches du Cyclotron, Universit\'{e} catholique de Louvain, B-1348 Louvain-la-Neuve, Belgium}

\date{\today}
\begin{abstract}
	
We have measured the \npg\ excitation function for energies in the range \ep\ = 155--524 keV. Fits of these data using  $R$-matrix theory yield a value for the S-factor at zero energy of $1.64\pm0.07$ ({\em stat}) $\pm0.15$ ({\em sys}) keV$\cdot$b, which is significantly smaller than the result of a previous direct measurement. The corresponding reduction in the stellar reaction rate for  \npg\ has a number of interesting consequences, including an impact on estimates for the age of the Galaxy derived from globular clusters.
\end{abstract}
\pacs{PACS: 25.40.Lw, 26.20.+f}
\keywords{\npg, CN cycle, $R$-matrix analysis}
\maketitle

Stars produce energy from the conversion of hydrogen into helium primarily by the
p-p chains and by the CN-cycle. The latter is the dominant energy source for stars somewhat more massive than the sun, but all stars will
produce energy via the CN cycle at the end of their main--sequence lifetimes, 
and while on the
red-giant branch. At the burning temperatures characteristic of these evolutionary
stages  (T $\approx$ 0.02--0.055 GK), the \npg\
reaction is the slowest CN reaction and thus it regulates the rate of nuclear energy
generation. The power liberated by the CN cycle and the amount of
helium produced has a direct connection to the luminosity observed
at the transition between the main sequence and the red--giant branch,
and on the luminosity of the horizontal branch. Both of these quantities
play a role in determining the ages of globular clusters \cite{cha1,cha2}. 
Also, since it helps to constrain the temperature and density profiles in the H-burning shell,
\npg\ will affect nucleosynthesis beyond the CN cycle during the red-giant stage. 

The thermonuclear reaction rate can be obtained from the astrophysical S-factor, defined
as 
\begin{equation}
S(E_{cm}) = E_{cm} \sigma(E_{cm}) \exp(2\pi\eta),
\label{sfact}
\end{equation}
where \ecm\ is the energy in the center of mass, $\sigma$(\ecm) is the reaction
cross section, and $\eta$ is the Sommerfeld parameter.
The accepted S-factor for the \npg\ reaction is based on the measurements of 
Schr\"oder {\em et al.}~\cite{Sch87} (hereafter Sch87) who quoted a zero-energy $S$-factor of $S(0)$ = $3.20\pm0.54$ keV$\cdot$b. About half of this S-factor results from transitions into the bound state at \ex\ = 6.79 MeV (J$^{\pi}$=$3/2^+$) in \ox. Most of the remainder arises from capture to the ground state (J$^{\pi}$=$1/2^-$), including a significant contribution from the tail of the 6.79-MeV state (corresponding to a subthreshold state at \ecm\ = 
-0.504 MeV).
However, uncertainty about the inferred width of the subthreshold state led to a recommendation of $S(0)$ = 1.5--4.5 keV$\cdot$b \cite{ade}, and a subsequent re-analysis of the data of Sch87 by Angulo and Descouvemont \cite{Ang01} reported S(0) = $1.77\pm0.20$ keV$\cdot$b. A
measurement of the lifetime of the 6.79-MeV state \cite{Ber01} strongly favored the lower range of values for $S(0)$. More recently, another re-analysis of the Sch87 data using measured asymptotic normalization coefficients (ANCs) as constraints produced a consistent result, $S(0)$ = $1.70\pm0.22$ keV$\cdot$b~\cite{Muk}. It should be noted that in the vicinity of the 0.259- and 0.985-MeV resonances, Sch87 present yield data rather than cross sections and that errors were made in their corrections for coincident summing~\cite{HPT}. Thus, a number of these points should not have been included in the analyses described above.
Nonetheless,  evidence favors a value for the $S$-factor that is  20--40\% smaller than previously thought for the temperatures of interest. Because of the importance of the \npg\ reaction in determining the power liberated by the CN
cycle, even a change of this magnitude impacts all of the issues mentioned above. In view of the importance of this reaction for stellar astrophysics, we have carried out a new measurement designed to more accurately determine the low-energy $S$-factor.

We measured an excitation function for the \npg\ reaction at the Laboratory for Experimental 
Nuclear Astrophysics (LENA), located at the Triangle Universities Nuclear 
Laboratory. A 1-MV Van de Graaff accelerator provided proton beams 
at laboratory energies between 155 and 524 keV, and with beam currents of 100 - 150 $\mu$A.
The beam entered the target chamber through a copper tube, extending to less than 1 cm from the surface of the target, which was cooled using chilled, de-ionized water. The copper tube was cooled by a LN$_{2}$
reservoir to trap potential target contaminants. In order to suppress the emission of secondary electrons from the target, permanent magnets were located at the end of the tube along with an electrode biased to -300 V . 
Targets were fabricated by implanting nitrogen ions into 0.5 mm-thick tantalum backings. The backings were prepared for implantation by etching in an acid solution \cite{ver53} to remove surface impurities that could give rise to beam-induced background. Implantation energies of 20, 40 and 110 keV were used to produce targets that were 5, 10 and 18-keV thick, respectively, as measured at the \ecm\ = 0.259-MeV resonance \cite{Ajz91}. Typically, their composition and thickness remained stable over accumulated doses of 20--25 C.  The condition of the target was checked periodically by measuring a yield curve for the 0.259-MeV resonance. The stoichiometry (Ta/N = 0.718$\pm$0.076) was measured via Rutherford backscattering and was consistent with published values~\cite{kei2}. From 32 independent measurements of the thick-target yield, we obtain \wg\ = 0.0135$\pm$0.0012 eV for the 0.259-MeV resonance, which is in good agreement with the previous value of 0.014$\pm$0.001 eV \cite{Bec82}. Our uncertainty is purely systematic and arises primarily from the uncertainty that we assign to the stopping powers (calculated using SRIM2000~\cite{zie}). By comparison, the statistical uncertainty (0.4\%) is negligible.  We have also measured the branching ratios for the decay of the 0.259-MeV resonance (listed in Table I). This measurement was carried out with the HPGe detector moved to a distance of 23 cm from the target, which reduced coincidence summing effects to a negligible level. It should be noted that our value for the ground-state transition differs significantly from the value of Sch87.

Gamma rays were detected using a 135\% HPGe detector placed at $\theta_{lab}$ = 0$^\circ$ and at a distance of 
9 mm from the target. The energy calibration and absolute photopeak efficiency were established 
using radioactive sources
and the decays from well-known resonances in the \npg, $^{26}{\rm Mg}(p,\gamma)^{27}{\rm Al}$ and 
$^{27}{\rm Al}(p,\gamma)^{28}{\rm Si}$ reactions. The total efficiency (needed for summing corrections) was calculated using MCNP~\cite{mcnp} and normalized to source data.
A 35.6-cm dia. x 40.6-cm long annulus of NaI(Tl) enclosed both the target and
Ge crystal. This detector geometry allowed us to record three types of events: Ge singles,
Ge-NaI coincidences, and Ge signals without a corresponding event in the NaI detector within 5 $\mu$s. In the latter mode, the NaI served as a
cosmic-ray veto while also suppressing events arising from $\gamma$-ray cascades, which proved useful for detecting the ground-state transition. Because of this fixed geometry, we could not measure angular distributions for the primary transitions. However, for our energy range, the primary transitions to the ground, 5.18 and 6.18-MeV states were calculated to be isotropic to better than 1\%. In addition, all secondary transitions were calculated to be isotropic. The only transition expected to have a significant angular distribution was to the 6.79-MeV state, which necessitated the use of the (isotropic) secondary transition alone to determine the cross section. The ratio of primary-to-secondary yields was consistent with the expected angular-distribution coefficient for an E1 transition, $a_{2}$=-1. Above the 0.259-MeV resonance, the secondary yields had to be corrected for the possibility that the incident proton could lose enough energy in the implanted target layer to undergo resonant capture into this state, which would produce some of the same secondaries. The signature of this effect was the presence of $\gamma$-rays from the de-excitation of the resonance, which also provided the means for removing their contribution to the secondaries. Because we used thin targets above the resonance, these corrections were small (7\% on average for the dominant transition to the 6.79-MeV state). 
\begin{table}[h]
\caption{\label{tab:table1}Branching ratios for the decay of the 0.259-MeV resonance}
\begin{ruledtabular}
\begin{tabular}{lccc}
 transition & this work & Sch87 \\ 
 \hline
ground state & $0.0170\pm0.0007$ & $0.035\pm0.005$\\
5.18 MeV & $0.173\pm0.002$ & $0.158\pm0.006$\\
6.18 MeV & $0.583\pm0.005$ & $0.575\pm0.004$\\
6.79 MeV & $0.227\pm0.003$ & $0.232\pm0.006$\\
\end{tabular}
\end{ruledtabular}
\end{table}
  
The effective center-of-mass energies, E$_{\rm eff}$, were derived from the energies of the primary $\gamma$-rays (corrected for Doppler shifts) as well as from the beam energies and measured target thicknesses. The second procedure was an iterative calculation, which first assumed a linear $S$-factor over the width of the target and then 
used the resulting S-factor to recalculate the result. These two techniques were found to yield consistent values. We also checked this procedure by measuring the S-factor with targets of different thicknesses, but at the same predicted values for E$_{\rm eff}$. These measurements produced S-factors that agreed to within experimental errors. The $\gamma$-ray yields at each energy were corrected for coincident summing, which was a significant effect. In fact, for energies below E$_{\rm eff}$ = 187 keV, the measured intensity at the energy of the ground state transition was entirely consistent with summing of cascades involving the higher-lying states. For the 6.79-MeV transition, the correction for summing-out was about 22\% These calculations included the (calculated) angular correlation between the primary and secondary $\gamma$-rays, which was a comparatively small effect since the product of the attenuation coefficients for the primary and secondary transitions amounted to about 0.15.

The three strongest transitions, in order of importance, are those to the 6.79-MeV state, the ground state (g.s.), and the 6.18-MeV state (J$^{\pi}$=$3/2^-$). 
These have been analyzed in the framework of the $R$-matrix model \cite{Lan58}, including the external contributions to radiative capture; details can be 
found in Ref.~\cite{Ang01}. These calculations have been performed with $R$-matrix radii of $a=3.5, 4, 4.5, 5, 5.5$, and 6 fm. 
To fit the 6.79-MeV transition, we have 
included the 0.259-MeV resonance ($J^{\pi} =1/2^+$) and the non-resonant contributions for
$J^{\pi} = 1/2^-, 3/2^-$ and $5/2^-$ channel spins, calculated from the ANC, $C$, of the final state.
The proton width $\Gamma_p$ of the resonance was left as free parameter, while
its gamma width was fixed for all fits to $\Gamma_{\gamma} = 9.2\pm0.1$ meV. This value 
was obtained from our measured resonance strength 
and the relevant branching ratio from Table I. The ANC was also
left as a free parameter. The results of the fits are given in Table II. Figure 1 shows the fits together with the present 
experimental data (the notation RC denotes "radiative capture"). The error bars reflect the statistical uncertainty for each point. The data for this transition from Sch87 (corrected for summing effects \cite{HPT}) are also shown in the figure for comparison,
but were not used in the fits. As can be seen in the figure, our new measurement extends the S-factor to lower energies, with higher accuracy.

\begin{table}[h]
\caption{\label{tab:table2}Results of the $R$-matrix fits for the 6.79 MeV transition}
\begin{ruledtabular}
\begin{tabular}{lccccc}
 $a$ & $C$ & $\gamma^2$  & $\Gamma_{\rm p}$  & $S(0)$ & $\chi^2/N$ \\ 
(fm) & (fm$^{-1/2}$) & (MeV) & (keV) & (keV$\cdot$b) & ($N=29$) \\ 
 \hline
 3.5  & 4.4 & 1.9 & 0.97 & 1.1 & 1.4 \\
 4  & 4.5 & 1.2 & 0.96 & 1.1 & 1.4 \\
 4.5  & 4.5 & 0.8 & 0.94 & 1.1 & 1.5 \\
  5  & 4.6 & 0.5 & 0.93 & 1.1 & 1.5 \\
 5.5  & 4.6 & 0.4 & 0.91 & 1.1 & 1.6 \\
 6  & 4.7 & 0.3 & 0.89 & 1.2 & 1.7 \\
\end{tabular}
\end{ruledtabular}
\end{table}

\begin{figure}
\includegraphics[clip=true,width=8.7cm,scale=1.0,angle=0]{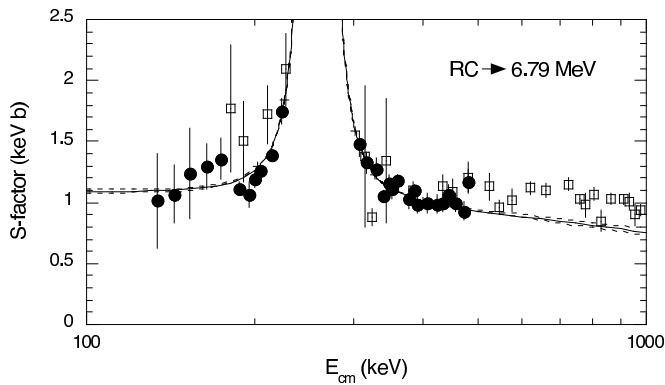}
\caption{$R$-matrix fits of the present capture data to the 6.79 MeV state (closed circles). The solid line shows the fit for $a=5$ fm; the lower (upper) dashed lines are for $a=4.5 (5.5)$ fm.
The corrected data from Sch87 [8] are also shown (open squares) but are not used in the fit.}
\end{figure}

The fits are weakly sensitive to the $R$-matrix radius $a$, as expected. The resulting values for $C$ and $\Gamma_{\rm p}$ are also not very sensitive to $a$. On the other hand, for a given $a$,
the $\chi^2$ surface is relatively flat, indicating that several sets of ($C$,$\Gamma_{\rm p}$) 
parameters are possible, but the variations are within the adopted uncertainty. 
We recommend $C = 4.5\pm0.1$ fm$^{-1/2}$ and $\Gamma_{\rm p} = 0.94\pm0.03$ keV.
The former value is in good agreement with 
an averaged value of the results from Mukhamedzanov et al. \cite{Muk} 
and Bertone et al. \cite{Ber02} ($C = 4.7\pm1.9$ fm$^{-1/2}$), while 
$\Gamma_{\rm p}$ is in very good agreement with the results of Sch87.
For the $S$-factor at zero energy, we have obtained 
$S(0) = 1.10\pm0.05$ keV$\cdot$b. The uncertainties were calculated from the $R$-matrix fits for $a=3.5-6$ fm that 
have $\chi^2 = \chi^2_{\rm min}+1$.

The $R$-matrix analysis of the g.s.~transition is
complicated by the presence of several contributing states. These involve the
E1 contributions from the $1/2^+$ resonance at 0.259 MeV, 
from the two $3/2^+$ resonances at 0.985 and 2.187 MeV, and from 
the $3/2^+$ subthreshold state at \ecm\ $= -0.504$ MeV. 
In order to account for higher-energy resonances, we have included a background pole ($\ell=1$)
at 4 MeV with J$^{\pi}$=$3/2^+$ and with $\Gamma_p$=4 MeV. 
The location and width of this state do not change S(0) by more than 15\%.
Because of this large number of parameters, it was necessary to reduce the number of free parameters in the fit. 
The proton and gamma widths for the 0.985- and 2.187-MeV resonances were taken from Ref.~\cite{Ang01}. We have also used $\Gamma_{\gamma} = 0.69\pm0.03$ meV for the 0.259-MeV resonance, obtained from our value for \wg\  and the branching ratio to the ground state. 
The ANC is taken from Table II (for each value of $a$), while the reduced width of the subthreshold state $\gamma^2$
is obtained from Eq. (14) of Ref.~\cite{Ang01} and is also given in Table II.
The number of free parameters is thus reduced to three: 
$\Gamma_{\gamma}^{\rm s}$ for the subthreshold state, $\Gamma_{\rm p}$ for the 
0.259-MeV resonance, and the gamma width of the 
background pole, $\Gamma_{\gamma}^{\rm bkg}$.
The results of the fits are given in Table III. The fit for $a=5$ fm is shown in Figure 2. For comparison, the data of Sch87,
corrected for summing effects \cite{HPT}, are also shown,
but they were not used in the fits. Note that these points are now restricted to \ecm\ $\geq$ 324 keV whereas the present data extend down to \ecm\ = 187 keV.
The proton width of the 0.259-MeV resonance, $\Gamma_{\rm p} = 1.3\pm0.1$ keV, 
is larger than the values of Table II obtained from the fits of the 6.79 MeV transition, 
but the error is larger here because of the more uncertain value for the branching 
ratio and the uncertainty associated with the background pole. For the subthreshold state, the value of $\Gamma_{\gamma}^{\rm s}$ strongly depends on 
the reduced width $\gamma^2$. In fact, the product $\Gamma_{\gamma}^{\rm s} \times \gamma^2$ does not depend strongly on the fit conditions \cite{Ang01}. Our values for $\Gamma_{\gamma}^{\rm s}$ are mostly higher than the 90\% confidence interval of 0.28--0.75 eV from Ref.~\cite{Ber01}. 
However, it is important to
note that the $S(0)$ value for the g.s. transition is not very sensitive to the fit parameters and we
have adopted $S(0) = 0.50 \pm 0.05$ keV$\cdot$b. While this value seems to be at odds with that of Ref.~\cite{Ber02}, the $S(0)$ quoted there was for direct capture only and did not include the interference effects of resonances, which are clearly important.\\

\begin{table}[h]
\caption{\label{tab:table3}Results of the $R$-matrix fits for the g.s.~transition}
\begin{ruledtabular}
\begin{tabular}{lccccc}
 $a$  &  $\Gamma_{\gamma}^{\rm s}$   & $\Gamma_{\rm p}$ & $\Gamma_{\gamma}^{\rm bkg}$ & $S(0)$  & $\chi^2/N$ \\ 
 (fm) &  (eV) & (keV) & (eV) & (keV$\cdot$b) & ($N=21$) \\ 
   \hline
 3.5  & 0.4 & 1.3 & 50 & 0.52 & 0.54 \\
 4  & 1.1 & 1.3 & 60 & 0.53 & 0.54 \\
 4.5  & 1.7 & 1.3 & 60 & 0.53 & 0.54 \\
  5  & 1.9 & 1.3 & 50 & 0.47 & 0.53 \\
 5.5  & 2.2 & 1.3 & 40 & 0.46 & 0.53 \\
 6  & 2.6 & 1.3 & 40 & 0.47 & 0.54 \\
\end{tabular}
\end{ruledtabular}
\end{table}

\begin{figure}[h!]
\includegraphics[clip=true,width=8.7cm,scale=1.0,angle=0]{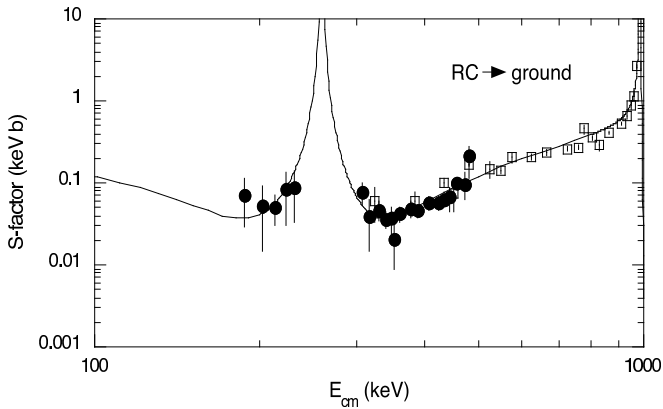}
\caption{$R$-matrix fit ($a=5$ fm) of the present capture data to the ground state (closed circles). 
The corrected data from Sch87 are also shown (open squares) but are not used for the fit.}
\end{figure}

Fits for the transition to the 6.18-MeV state have been performed, including 
E1 contributions from the  0.259-MeV resonance, the subthreshold 
state, and a background pole.
Figure 3 shows two fits with different values of $\Gamma_{\gamma}^{\rm s}$ for the 
subthreshold state and different parameters for the background pole. 
Both fits give very similar (and large) $\chi^2$ values. 
We have adopted an intermediate value of $S(0) = 0.04\pm0.01$ keV$\cdot$b. Although the $S$-factor for this transition is rather uncertain, it has a small effect on the total $S$-factor.
In Figure 3, we also show the M1 contribution calculated using the parameters given by  Nelson {\em et al.}~\cite{Nel03}. In contrast to their conclusion, we find this contribution to be negligible at low energies.

The present extrapolated $S$-factor at zero energy is compared in Table V with previous results. In addition to the statistical uncertainty quoted there, we also estimate systematic uncertainties associated with the measurement of total charge ($\approx$ 2.5\%), photopeak efficiency ($\approx$ 3\%), effective energy ($\approx$ 3\% at the lowest energies) and target composition ($\approx$ 8\%). If the systematic uncertainties are combined in quadrature, then they amount to an overall uncertainty of 9.4\% at the lowest energies. Consequently, our result is $S(0)$ = $1.64\pm0.07$ ({\em stat}) $\pm0.15$ ({\em sys}) keV$\cdot$b, which is 51\% of the value of Sch87. Although our result for $S(0)$ agrees with a reanalysis of the Sch87 data.~\cite{Ang01}, the values for all 3 transitions are quite different (and thus is the energy dependence of the S-factor). Note also that that our findings are consistent with another recent measurement of \npg\ at low energies~\cite{HPT}.
To further improve on the accuracy of $S(0)$, new data are needed at lower energies, but also at energies above 0.5 MeV.

We have explored one of the consequences of this result by calculating the evolution of a star with a mass of 0.8 times the mass of the sun. The composition was appropriate for the galactic halo, with an overall metalicity Z = 1.7 $\times$ 10$^{-4}$ (or 1\% of solar). With the present rate for \npg\ we find that the age at the main-sequence turnoff is 0.8 Gy older than that with the previous rate. The implication is that globular-cluster ages will have to be revised upwards. Further investigation of this issue and of other implications arising from this work are in progress.

\begin{figure}
\includegraphics[clip=true,width=8.7cm,scale=1.0,angle=0]{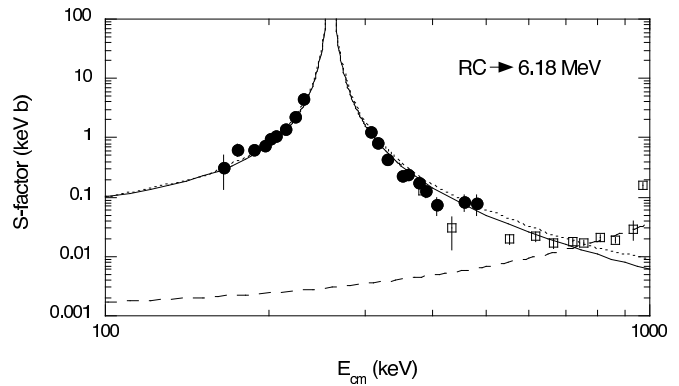}
\caption{$R$-matrix fit ($a=5$ fm) of the capture data to the 6.18 MeV state (closed circles). 
The data from Sch87 are also shown (open squares) but are not used for the fit.
The solid (dotted) line is the fit with a background pole at 3.0 MeV (5 MeV); the dashed curve represents an M1 contribution (see text). }
\end{figure}

\begin{table}[h]
\caption{\label{tab:table3}Results of the $R$-matrix fits for 6.18 MeV state for $a=5.5$ fm}
\begin{ruledtabular}
\begin{tabular}{cccccc}
$\Gamma_{\gamma}^{\rm s}$  & $E_{\rm r}^{\rm bkg}$ & $\Gamma_{\rm p}^{\rm bkg}$ & $\Gamma_{\gamma}^{\rm bkg}$ & $S(0)$  & $\chi^2/N$ \\
 (meV) & (MeV) &  (MeV) & (eV) & (keV$\cdot$b) & ($N=19$) \\ \hline
5.0 & 3 & 1.5 & 1.0 & 0.04 & 4.9 \\
0.5 & 5 & 1.5 & 0.1 & 0.03 & 5.8 \\
\end{tabular}
\end{ruledtabular}
\end{table}

\begin{table}[h]
\caption{\label{tab:table4}Summary of $S(0)$ values (in units of keV$\cdot$b)}
\begin{ruledtabular}
\begin{tabular}{lccc}
Transition & Ref.~\cite{Sch87}  & Ref.~\cite{Ang01} &  Present \\
(MeV)      &       &  &   \\
\hline 
RC $\rightarrow$ 0       & $1.55\pm0.34$  & $0.08^{+0.13}_{-0.06}$ & $0.50\pm0.05$\\
RC $\rightarrow$ 6.18    & $0.14\pm0.05$  & $0.06^{+0.01}_{-0.02}$ & $0.04\pm0.01$\\
RC $\rightarrow$ 6.79    & $1.41\pm0.02$  & $1.63\pm0.17$          & $1.10\pm0.05$ \\
Total                    & $3.20\pm0.54$  & $1.77\pm0.20$          & $1.64 \pm 0.07$ \\
\end{tabular}
\end{ruledtabular}
\end{table}

\begin{acknowledgments}
This work was supported in part by the U.S. Department of Energy 
under Contract No. DE-FG02-97ER41041 and in part by the Belgian PAI program
P5/07 on Interuniversity Attraction Poles.
We would like to thank P. Descouvemont for the $R$-matrix code and H.-P. Trautvetter for fruitful discussions.
\end{acknowledgments}


\begin{references}

\bibitem{cha1}B. Chaboyer {\em et al.}, Science \textbf{271}, 957 (1996).
\bibitem{cha2}B. Chaboyer{\em et al.} Astrophys. J. \textbf{494}, 96 (1998).
\bibitem{Sch87} U. Schr\"oder {\sl et al.}, Nucl. Phys. A\textbf{467}, 240 (1987).
\bibitem{ade}E.G. Adelberger {\em et al.}, Rev. Mod. Phys. \textbf{70}, 1265 (1998).
\bibitem{Ang01} C. Angulo and P. Descouvemont, Nucl. Phys. A\textbf{690}, 755 (2001).
\bibitem{Ber01} P.F.~Bertone {\em et al.}, Phys. Rev. Lett. \textbf{87}, 152501 (2001).
\bibitem{Muk}A.M. Mukhamedzhanov {\em et al.}, Phys. Rev. C\textbf{67}, 065804 (2003).
\bibitem{HPT} H.-P. Trautvetter for the LUNA collaboration, private communication.
\bibitem{ver53}D.A. Vermilyea, Acta Metallurgica \textbf{1}, 282 (1953).
\bibitem{Ajz91} F. Ajzenberg-Selove, Nucl. Phys. A\textbf{523} (1991) 1.
\bibitem{kei2}J. Keinonen and A. Anttila, Nucl. Instr. Meth. {\bf 160}, 211 (1979).
\bibitem{Bec82}H.W. Becker {\em et al.}, Z. Phys. A\textbf{305}, 319 (1982).
\bibitem{zie}J.F. Ziegler and J.P. Biersack, program SRIM2000 (2000), unpublished.
\bibitem{mcnp}Diagnostic Applications Group, LANL, program MCNP4C (2000), unpublished. 
\bibitem{Lan58} A.M. Lane and R.G. Thomas, Rev. Mod. Phys. \textbf{30}, 257 (1958).
\bibitem{Ber02} P.F. Bertone {\em et al.}, Phys. Rev. C\textbf{66}, 055804 (2002).
\bibitem{Nel03} S.O. Nelson {\em et al.}, Phys. Rev. C\textbf{68}, 065804 (2003).


	
\end{references}
\end{document}